\documentclass[3p,twocolumn]{elsarticle}

\usepackage{lineno,amsmath,amssymb,upgreek}
\usepackage[hidelinks]{hyperref}
\modulolinenumbers[5]
\usepackage[dvipsnames,table]{xcolor}
\usepackage{soul}
\usepackage{booktabs}

\colorlet{hy}{Goldenrod!50}
\newcommand{\og}{\cellcolor{OliveGreen!20}}
\newcommand{\yl}{\cellcolor{Goldenrod!30}}

\usepackage{xparse}
\NewDocumentCommand{\todo}{s m}{%
    \sethlcolor{hy}%
    \IfBooleanTF{#1}{\textcolor{gray}{\sout{#2}}}{\hl{#2}}%
}

\makeatletter
\def\ps@pprintTitle{%
 \let\@oddhead\@empty
 \let\@evenhead\@empty
 \def\@oddfoot{}%
 \let\@evenfoot\@oddfoot}
\makeatother

\biboptions{authoryear}

\begin{document}

\begin{frontmatter}

\title{The threshold at which a meteor shower becomes hazardous to spacecraft}

\author{Althea V.\ Moorhead\footnotemark[1]}
\author{William J.\ Cooke}
\address{NASA Meteoroid Environment Office, Marshall Space Flight Center, Huntsville, AL 35812, USA}

\author{Peter G.\ Brown}
\author{Margaret D.\ Campbell-Brown}
\address{Department of Physics and Astronomy, University of Western Ontario, London, Ontario N6A 3K7, Canada}

\begin{abstract}
Although the risk posed to spacecraft due to meteoroid impacts is dominated by sporadic meteoroids, meteor showers can raise this risk for short periods of time. NASA's Meteoroid Environment Office issues meteor shower forecasts that describe these periods of elevated risk, primarily for the purpose of helping plan extravehicular activities. These forecasts are constructed using a list of meteor shower parameters that has evolved over time to include newly discovered showers and incorporate improved measurements of their characteristics. However, at this point in time, more than a thousand meteor showers have been reported by researchers, many of which are extremely minor, are unconfirmed, or lack measurements of critical parameters. Thus, a comprehensive approach is no longer feasible. In this report we present a quantitative criterion for a potentially hazardous meteor shower and apply this criterion to the list of established meteor showers in order to determine which showers should be included in our annual forecasts.
\end{abstract}


\end{frontmatter}

\section{Introduction}

Meteoroid impacts can damage spacecraft and have been linked to a number of reported anomalies \citep[see][for an overview]{handbook}. Most hazardous meteoroids belong to the so-called ``sporadic complex,'' which is present throughout the year. The constant risk of sporadic meteoroid strikes requires spacecraft engineers to add shielding or redundancies to spacecraft in order to protect against a loss of mission. However, meteor showers can elevate the risk for short periods of time, and the risk they pose can be mitigated operationally by phasing an orbit, turning a less vulnerable side to the shower radiant, or delaying an extravehicular activity
\citep[also known as an EVA or ``spacewalk''; see][for an overview of mitigation strategies]{peterson99,drolshagen19}.

\footnotetext[1]{althea.moorhead@nasa.gov}

All mitigation procedures have a cost, whether that cost is monetary or one of opportunity. Quantitative models of the meteoroid environment are needed in order to weigh these costs against the risk of damage. Without a quantitative prediction, one cannot assess the risk and may be forced to abort or delay a mission. For instance, the launch of space shuttle Discovery in 1993 on mission STS-51 was delayed due to predictions of a Perseid outburst of unknown strength \citep{broad93}. A shuttle launch had never before been delayed due to a meteor shower and, thanks to advances in meteor shower modeling and forecasting, none have been delayed by a meteor shower since.

Accurate meteor shower forecasting depends on several factors. First and foremost, it relies on accurate and detailed observations of meteor showers that span many years (and, ideally, decades). Such observations are needed to calibrate models and predict future activity. Second, it depends on our ability to recreate a shower's past activity with modeling. If the parent body of the stream is unknown, or if its past motion and activity are poorly understood, it may be difficult-to-impossible to predict future behavior even when recent, high-quality observations of the meteor shower are available. Finally, we must be able to correctly translate meteoroid stream models into quantities that are useful for spacecraft risk assessments. 

It is therefore important to focus our efforts on those showers that have the potential to be dangerous to spacecraft.
We define a potentially hazardous meteor shower as a meteor shower for which the flux of particles carrying a kinetic energy of 105~J or greater at the time of its peak activity is at least 5\% of the sporadic meteoroid flux on a flat plate facing the unobscured shower radiant at low altitudes above the Earth in any year. In section~\ref{sec:crit} we justify each component of this definition. In section~\ref{sec:methods}, we discuss our methods for selecting showers to assess and compute the minimum activity level needed for showers to meet our definition of a potentially hazardous shower. Finally, in section~\ref{sec:list}, we determine which showers meet our criterion and revise the list of showers we forecast accordingly.

\section{Definition justification}
\label{sec:crit}

We opt to define a threshold rather than quantify the damage rate associated with a meteor shower \citep[as][does]{ma06}. We take this approach because the mass distributions of many meteor showers are unknown. A threshold permits us to ask whether a given meteor shower is capable of posing a threat for \emph{any} reasonable choice of population index. The population index is a quantity that measures the slope of the meteoroid magnitude distribution; meteor magnitude is often used as a proxy for meteoroid mass \citep[e.g.,][]{jvb67,verniani73,cb16}. In this section we deconstruct our definition of a potentially hazardous meteor shower and justify each element in turn.

\subsection{Kinetic energy threshold}

Meteoroids are capable of producing a variety of effects such as penetrating surfaces, severing wires, transferring momentum, and triggering electrostatic discharges by producing plasma \citep{peterson99,garrett13,drolshagen19}. These effects have different dependences on the meteoroid's mass, density, size, impact speed, and impact angle; even a single effect, such as penetration depth, will have a different relationship to meteoroid properties for different target materials. As a result, it is not possible to set a single ``hazard'' threshold that applies to all meteoroid impact effects.
\citep[][attempts to address this using a hazard index based on a weighted average of different effect enhancements, but includes non-risk-related terms such as the mass-limited flux enhancement and acknowledges that the weights are arbitrary.]{wu06} 

We note, however, that risk assessments most frequently focus on the probability of no penetration (PNP). The particle size, speed, density, and impact angle that are capable of penetrating a spacecraft component are computed using a ballistic limit equation (BLE) that is appropriate for the component material. The most common BLEs, such as the modified Cour-Palais BLE \citep{hayashida91}, are nearly proportional to the kinetic energy of the particle (or rather, the kinetic energy raised to the $2/3$ power). Thus, our meteor shower forecasts use kinetic energy because it is a simple limiting quantity and because kinetic-energy-limited flux enhancements will resemble damage-limited flux enhancements. The use of kinetic energy neglects the typical dependence of BLEs on impact angle, but this can be accounted for, if desired, by applying a factor of two \citep{moorhead19b}.

Both meteoroid streams and the sporadic meteoroid complex have size distributions that resemble an exponential; that is, we assume that the number of shower particles above a given mass threshold $m$ can be expressed as
\begin{align}
N_{> m} &\propto m^{-(s-1)} \label{eq:index}
\end{align}
where the exponent $s$ is known as the (differential) mass index. 
It is related to the population index, $r$, as follows:
\begin{align}
    r &= 10^{(s-1)/2.3} \, .
\end{align}
The quantity 2.3 is derived from the mass-magnitude relation of \cite{verniani73}; other authors may use a value of 2.5, which assumes that peak brightness is directly proportional to the initial particle mass.

\begin{figure}
    \centering
    \includegraphics{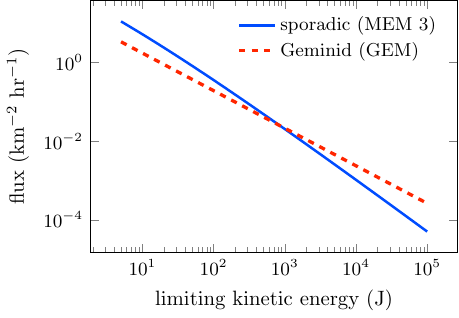}
	\caption[Flux as a function of limiting energy]{Sporadic and sample shower (GEM) flux near the top of the Earth's atmosphere as a function of limiting kinetic energy.}
\label{fig:fke}
\end{figure}

Shower mass indices typically range from values of about 1.7 \cite[for, e.g., the Quadrantids;][]{pokorny16} to 2.2 \cite[for the Piscis Austrinids;][]{jenniskens94}. In contrast, the sporadic mass index is believed to be 2.3 for large sporadic particles. Thus, the relative strength of shower and sporadic fluxes will depend on the limiting mass or energy in question. As an example, Fig.~\ref{fig:fke} compares the Geminid (GEM) flux to the sporadic flux predicted by version~3 of the Meteoroid Engineering Model \citep[MEM~3;][]{moorhead20}.

Selecting an appropriate limiting energy for assessing shower significance is a balancing act. When the limiting energy is low, showers constitute a smaller fraction of the risk, but particles may be too small to cause any real damage. If we set the energy limit too low, we might therefore underestimate the importance of showers. When the limiting energy is high, showers are a larger fraction of the risk, but the flux of both showers and sporadic particles are low. If we set the limit too high, we might overestimate the importance of showers. It is important to select an energy threshold that corresponds to the spacecraft damage threshold as closely as possible, although this will of course vary from spacecraft to spacecraft and from material to material.

The Meteoroid Environment Office (MEO) typically issues forecasts that report fluxes and flux enhancement factors to four limiting kinetic energies: 6.7~J, 105~J, 2.8~kJ, and 105~kJ \citep{moorhead19b}. These limits correspond roughly to particles that are capable of penetrating a spacesuit, ``delicate'' spacecraft components, ``robust'' spacecraft components, and causing mission-ending damage, respectively. We discard the larger two sizes as candidate thresholds for this analysis. These higher damage thresholds are more pertinent to assessing the risk of damage to the spacecraft body, which will be continuously exposed to the environment with little-to-no opportunity for the kind of temporal mitigation that the shower forecasts allow. The lower two energy thresholds are more relevant if one wants to consider delaying a spacewalk or turning a spacecraft to shield delicate components from an active meteor shower. Thus, we select the 105~J threshold as that corresponding to the largest meteoroid sizes for which we can temporarily shield an astronaut or one component of a spacecraft.

\subsection{Peak activity}

We choose to assess a shower's contribution to the overall meteoroid flux at the moment of the shower's peak activity. This is quite different from a shower's total contribution to the meteoroid flux, as meteor shower activity decays exponentially with the temporal offset from the peak \citep{jenniskens94}, while sporadic activity remains relatively constant. Thus, a shower's contribution averaged over any time period will necessarily be less than its contribution at the time of the peak. It is already known that showers are responsible for a small fraction of the meteoroid flux on average \citep{moorhead19b}, and we therefore focus on quantifying the maximum enhancement in order to support assessments of the benefits of operational mitigation.

On the other hand, it's worthwhile to note that even a large outburst could pose a minimal risk to spacecraft if it is extremely brief. For instance, a doubling of the hazardous meteoroid flux that lasts only a few minutes poses no more risk to an astronaut or asset than a few additional minutes of exposure to the sporadic meteoroid environment. Before a new shower is incorporated into our forecasts, we will assess whether the risk is increased by 5\% for at least one hour. However, because duration information is not available for all showers, we will neglect duration for the time being and instead consider the peak flux or zenithal hourly rate (ZHR); the ZHR is the number of meteors a visual observer sees under ideal observing conditions and when the radiant is at local zenith.

\subsection{Enhancement percentage}

The uncertainty in the sporadic meteoroid flux is approximately a factor of two to three \citep{iadc,moorhead20}. Furthermore, unmodeled seasonal trends in the sporadic sources can produce variations in the sporadic flux of up to 40\% \citep{cb06}, although these variations occur over timescales (months) that are not conducive to operational mitigation. Our required 5\% enhancement over the sporadic background is small compared to the uncertainty and observed variations in the overall flux and errs on the side of shower inclusion.

\begin{figure}
    \centering
    \includegraphics{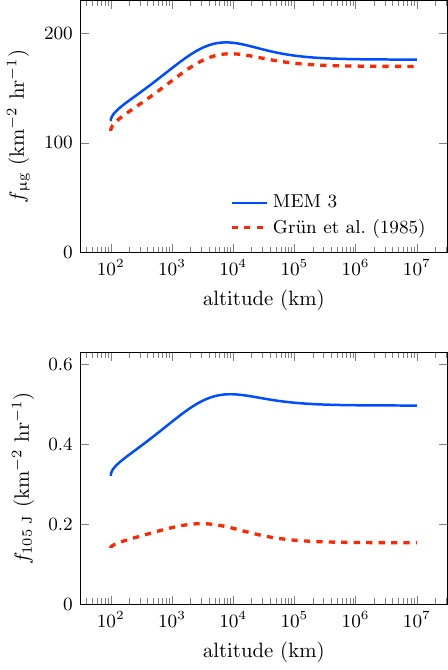}
	\caption[Comparison of fluxes from MEM~3 and \cite{grun85}]{The flux of meteoroids with masses greater than 1~$\upmu$g (top) or energies greater than 105~J (bottom) according to MEM~3 (solid blue line) or \citet[][dashed red line]{grun85} as a function of altitude.
}
	\label{fig:memvgrun}
\end{figure}

The selection of a relatively low 5\% enhancement also leaves room for any potential underestimates of shower activity or overestimates of sporadic activity. For instance, MEO forecasts have historically compared shower fluxes to the \cite{grun85} sporadic flux. However, as discussed in section~\ref{sec:spoflux}, the kinetic-energy-limited \citeauthor{grun85}\ flux is approximately a factor of two lower than the MEM~3 flux (see Fig.~\ref{fig:memvgrun}). A shower that is capable of producing a a 5\% enhancement over the MEM~3 flux then produces a 10\% enhancement over the \citeauthor{grun85}\ flux. However, this is still well within the uncertainty in the sporadic flux.

\subsection{Flat plate as target}

If a flat plate directly faces a shower radiant, the ratio of the shower flux to the sporadic flux is four times higher on that plate than it is averaged over the entirety of the spacecraft \citep{moorhead19b}. The MEO's forecasts typically use this flat-plate ratio in order to better describe the increase in risk experienced by a spacecraft element that is directly facing an unobscured radiant. We will also adopt this approach here.

Our enhancement threshold therefore corresponds to a 5\% enhancement of the kinetic-energy-limited meteoroid flux incident on a flat plate that is directly facing the unobscured shower radiant. This is equivalent to a 1.25\% enhancement of the flux encountered by a randomly tumbling object (i.e., one that has no special orientation relative to the meteoroid environment) on all surfaces.

In \cite{moorhead19b}, we noted that the damage done by a meteoroid impact is often assumed to depend on the impact angle, where grazing impacts are less likely to penetrate spacecraft surfaces. Thus, the penetration-limited flux enhancement can be greater than the kinetic-energy-limited flux enhancement for a component that faces the shower radiant by roughly a factor of two. Thus, our required 5\% enhancement in the energy-limited flux corresponds approximately to a 10\% enhancement in the penetration-limited meteoroid flux.

\subsection{Altitude}

While our 5\% threshold may seem straightforward, it is complicated by the fact that sporadic and shower fluxes vary with altitude. At low altitudes, the Earth's gravity accelerates meteors and increases their flux through a phenomenon known as gravitational focusing \citep[see, e.g.,][]{kessler72,jones07}. Slow meteors experience a proportionally greater acceleration and flux enhancement; when averaged over the globe, the enhancement in the mass-limited flux at a given altitude $h$ is
\begin{align}
    \frac{f_h}{f_g} &= 1 + \frac{v_\mathrm{esc}^2(h)}{v_g^2}
\end{align}
where $f_h$ is the meteoroid flux at altitude $h$, $f_g$ is the flux ``at infinity'' (i.e., at arbitrarily high altitudes), $v_\mathrm{esc}^2(h)$ is the escape velocity at altitude $h$, and $v_g$ is the speed of the meteoroid at infinity. While it might be more intuitive to use the symbol ``$\infty$'' to denote ``at infinity,'' we adopt the more commonly used subscript $g$ in order to avoid confusion.

Gravitational focusing at low altitudes will be counteracted to some degree by planetary shielding, in which the Earth physically shields the spacecraft from a portion of the environment. This reduces the average sporadic flux onto a low-orbiting spacecraft, but does not reduce the shower flux on a flat plate that faces the unobscured shower radiant. Thus, planetary shielding will tend to increase the relative importance of meteor showers at low altitudes.

\begin{figure}
    \centering
	\includegraphics{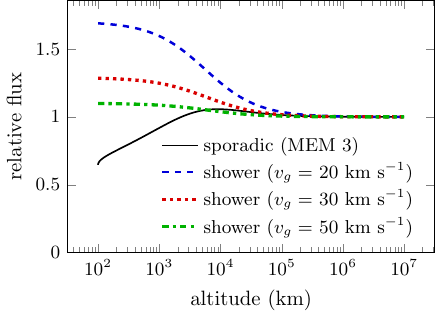}
	\caption[The dependence of meteoroid flux on altitude]{The ratio of the kinetic-energy-limited meteoroid flux, as modeled using MEM~3 or the methodology of \cite{moorhead19b}, at a given altitude to that in interplanetary space.}
	\label{fig:alt}
\end{figure}

In Fig.~\ref{fig:alt}, we plot the variation in flux with respect to altitude for both the sporadic meteoroid speed distribution (as modeled by MEM~3) and three sample shower speeds. We do not plot the flux itself, but rather the flux of a population at a given altitude relative to its flux in interplanetary space. We find that at intermediate altitudes ($\sim 10\,000$~km) above the Earth, the sporadic flux is a little higher due to gravitational focusing. However, at low altitudes, planetary shielding is the dominant effect and the sporadic flux is diminished relative to its interplanetary level.

The flux of an unobscured meteor shower, on the other hand, will be at its highest level at low altitudes. Thus, the largest enhancements due to meteor showers occur at the top of the Earth's atmosphere, regardless of shower speed. We can therefore determine whether a shower is significant at any altitude by quantifying its significance at an altitude of 100~km. We do not consider altitudes below 100~km, because this is the point at which the atmosphere becomes dense enough to ablate meteoroids, effectively removing them as a potential hazard.

\subsubsection{Gravitationally focused streams}

If a spacecraft orbiting the Earth at high altitudes passes through a point that lies directly opposite the planet from the shower's apparent point of origin, the Earth can act as a lens and produce a large enhancement of the shower's flux at the spacecraft's location. A spacecraft's orbital motion, however, ensures that it cannot remain near this location for long. Additionally, \cite{moorhead20b} found that when the gravitational focusing is modeled realistically, the enhancement is less than a factor of 10, although the exact factor depends on the shower's speed and the dispersion within the stream. Furthermore, this effect cannot occur at low altitudes due to planetary shielding, and thus we do not incorporate it into this analysis.

\subsubsection{Restriction to near-Earth space}

The streams responsible for creating meteor showers arise when sublimating gases carry particles away from active comets. (Asteroids have also been linked to some meteoroid streams but are not the main source.) These particles spread over time, but remain on orbits somewhat similar to that of their parent comet. They therefore remain organized in a stream of particles; meteor showers occur when the Earth passes through the stream.

It is not possible, however, to predict which comets produce substantial streams. Comets have variable behavior and may or may not produce significant numbers of meteoroids at any given point in time. Thus, one cannot simply survey the list of known comets and predict meteor shower activity by selecting those comets that pass near the Earth or are capable of producing debris that would pass near the Earth. This approach can be used to identify the \emph{possibility} of a new meteor shower \citep[see, e.g.,][]{lyytinen03}; if the comet's recent activity has been observed, these predictions can become more reliable \citep{ye15}. However, the translation of comet activity to meteor shower activity is extremely difficult and the meteor rate generally cannot be predicted without calibrating one's model to match previous shower activity.

Since we rely on past meteor shower observations to calibrate stream models, it is virtually impossible to forecast meteor shower activity on other planets. We have almost no data pertaining to meteor showers on other planets; the only exception is the Taurid meteoroid stream, whose activity at Mercury may produce enhancements in the planet's calcium exosphere \citep{killen15,christou15,christou24}. A link between meteor showers and methane plumes on Mars has been proposed \citep{fries15} but the correlation is debated \citep{roos16}. Given this context, the MEO refrains from issuing meteor shower forecasts for locations outside near-Earth space.

\subsection{Year of activity}

Many showers vary in their level of activity from year to year; some showers have also been shown to vary in population index and timing \citep[see][for an example]{koten20}. Some showers, such as the Leonids, are capable of producing rare but enormous outbursts of activity that increase the hazardous flux by factors of hundreds or even thousands. Therefore, it is important to consider not only the typical activity level of a shower, but the highest level of activity it is capable of producing in any year.

\section{Methods}
\label{sec:methods}

\subsection{Shower selection and activity data}
\label{sec:rates}

The International Astronomical Union (IAU) Meteor Data Center (MDC)\footnote{https://www.ta3.sk/IAUC22DB/MDC2007/index.php} maintains a list of meteor showers and their basic characteristics. There is no minimum activity level required for inclusion; any published (or soon-to-be-published) meteor shower is eligible for inclusion in the working list. A smaller set of showers are designated as ``established'' after they have been subjected to a verification process and their proposed names accepted at the IAU General Assembly \citep{jopek17}.

The full working list contains over 1000 showers, most of which lack a usable estimate of their activity level. We take the view that if the MDC has not yet been able to verify these showers, reliable estimates of their peak ZHR values or population indices likely do not exist in the literature. Therefore, we will generally restrict our analysis to the 110 established showers. There are six exceptions: we include the 
ARD and 
CAM showers in our list of outbursting showers, and we include the 
AUM,
LDR,
LLY, and
NID showers because we have measurements of their peak flux \citep{profilesreport}.

\subsubsection{Established showers}

The MDC provides basic timing and orbital information for meteor showers, including geocentric speed and peak solar longitude. However, it does not list population index or ZHR. We therefore supplement the MDC data with ZHR and population index estimates supplied by the International Meteor Organization (IMO) meteor shower calendar \citep[e.g.,][]{cal23}, an older survey of visual observations \citep{jenniskens94}, and a list of shower parameters on the American Meteor Society (AMS) website.\footnote{https://www.amsmeteors.org/meteor-showers/2020-meteor-shower-list/} 

For those established showers which were not represented in either the IMO calendar, the \cite{jenniskens94} survey, or the AMS website, we performed a cursory search of the literature. For example, we have no activity measures for the kappa Serpentid meteor shower, and so entered ``KSE ZHR'' and ``serpentid ZHR'' as search terms in both the SAO/NASA Astrophysics Data System and Google search. 
This approach yielded ZHR and/or population index measurements or estimates for the 
AAN \citep{molau09feb}, 
BTA \citep{dewsnap21}, 
DSX \citep{kipreos22}, 
NZC \citep{holman12}, 
NCC \citep{sokolova22}, and 
ZPE \citep{dewsnap21} showers. However, this search was by no means exhaustive and additional measures may exist in the literature.

\begin{figure}
    \centering
    \includegraphics{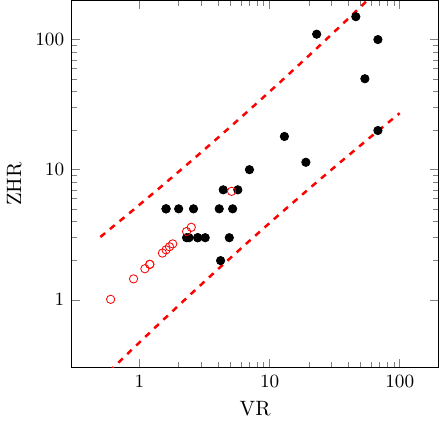}
	\caption{ZHR vs.\ \protect\cite{vr} video rate (VR) for those showers for which both values are available (black dots). We have fit a simple linear regression model to the logarithm of both quantities and used this model to estimate the ZHR for 12 additional showers (red open circles). The boundaries of the 90\% prediction interval are marked by red dashed lines.}
	\label{fig:vz}
\end{figure}

As a last resort, we used the \cite{vr} video rates as a proxy for ZHR. These video rates do not directly measure, but are instead correlated with, ZHR (see Fig.~\ref{fig:vz}). For those showers with both ZHR and video rate measurements, we fit a simple linear regression model to the logarithm of ZHR and video rate. We used this model to generate ZHR estimates for 12 additional established showers (AUD, AHY, CAN, DAD, HVI, JLE, NUE, OCU, OER, PSU, SCC, XCB). We considered using the upper bound of the 90\% prediction interval as an upper estimate of the ZHR of these showers, but this resulted in implausibly large ZHR estimates for showers that are rarely seen (such as a ZHR of over 20 for the NUEs).

\subsubsection{Outbursting showers}

Some showers are highly variable and produce orders of magnitude more meteors in certain years. Most of these highly variable showers are included in the MDC's list of established showers, including the 
AMO,
AND,
AUR,
BHY,
DRA,
JBO,
LEO,
LYR,
PHO, and 
PPU showers.

Two outbursting showers that are not in the list of established showers are the CAMs and ARDs. The CAM shower exhibited an outburst in May 2014 that, while visually underwhelming, appeared to be rich in small particles \citep{cb16} and likely produced a significant enhancement in the meteoroid flux. The ARDs produced an outburst in 2021 that was also rich in small particles and peaked at an estimated ZHR of 80 \citep{jenniskens21}.
Thus, we also include the CAMs and ARDs in our list of showers of interest.

Completely new meteor showers can sometimes be predicted by modeling the emission of particles from comets whose orbits pass near the Earth. In fact, both the CAMs and ARDs were new meteor showers that were successfully predicted by modelers \citep{jenniskens06,ye15}. While it is important to consider such showers in our forecasts, we will exclude theorized future showers from this work.

\subsubsection{Radar showers}
\label{sec:radar}

Meteoroids are equally capable of entering the Earth's atmosphere during the day or during the night, but, aside from the rare superbolide, the light they produce can only be detected at night. As a result, there is a bias against the detection of daytime meteor showers that can only be avoided through the use of meteor radar arrays, which detect ionization instead of light.

We recently completed a study in which we characterized the activity profiles of 38 showers using 20 years of single-station radar flux measurements from the Canadian Meteor Orbit Radar \citep[CMOR;][]{cb04,profilesreport}. We assumed that the observed activity rate in each case was a linear combination of the shower's activity and contamination by the sporadic background. This allowed us to subtract the sporadic background in order to better estimate the peak shower flux. 

The \cite{profilesreport} study relied primarily on flux data generated from an experimental branch of the CMOR flux pipeline that has better shower masking (to remove competing showers), but also differs from the operational pipeline in other aspects such as the assumed meteor trail length. However, the study also compared these flux measurements with those generated by the operational pipeline and with those calculated from 38~MHz observations (the default choice is 29~MHz). Based on this comparison, we estimated the systematic uncertainty to be approximately a factor of two for the fastest showers, and potentially much larger for low-speed showers. It may be as large as a factor of twenty for in-atmosphere speeds of 20~km~s$^{-1}$.

\begin{figure}
    \centering
    \includegraphics{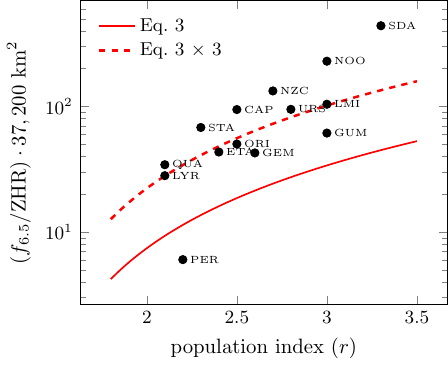}
	\caption{The ratio of the peak, magnitude-6.5-limited radar flux to ZHR for 15 showers, scaled by the maximum visual collecting area of 37\,200~km$^2$. The expected relationship (given by Eq.~\ref{eq:f65zhr}) is shown as a solid red line. Overall, the radar fluxes appear to be about three times the value one would expect from the ZHR values.}
	\label{fig:f65zhr}
\end{figure}

Here, we compare the nominal peak fluxes from \cite{profilesreport} with ZHR estimates for the same showers. CMOR fluxes are quoted to a constant limiting magnitude of +6.5; the magnitude-6.5-limited flux should in theory be proportional to the ZHR as follows \citep{koschack90a}:
\begin{align}
    \frac{f_{6.5}}{\text{ZHR}/37\,200~\text{km}^2} &= (13.1 r - 16.5)(r-1.3)^{0.748}
    \label{eq:f65zhr}
\end{align}
We have radar flux measurements, ZHR estimates, and $r$ estimates for fifteen showers in our master list. Fig.~\ref{fig:f65zhr} compares the observed flux-to-ZHR ratio with the expected ratio. With the exception of the Perseids (PER), fluxes are about three times as high as predicted from the ZHR values using Eq.~\ref{eq:f65zhr}. 

Because a factor of three is within two standard deviations of our nominal flux estimates at all speeds, we have opted to adjust the radar fluxes downward by this factor. We have made this choice so as not to artificially inflate the importance of showers whose flux has been measured by radar relative to those that have not. Note that a similar normalizing procedure between visual, video and radar meteoroid fluxes has been used in other studies \citep{egal20,egal22} which underscores the challenges in comparing absolute meteoroid fluxes across different observing techniques.

\subsection{Energy-limited sporadic flux}
\label{sec:spoflux}

Because we assess significance in terms of the fractional enhancement over the sporadic flux, we must first determine the flux of sporadic meteoroids with a kinetic energy above our chosen threshold. The shower forecasting algorithm described in \cite{moorhead19b} measures shower activity against a simple model of the meteoroid flux that is based on \cite{grun85}. This model takes the form of an analytic equation that describes the meteoroid flux as a function of limiting mass \citep[see Eq.~A3 of][]{grun85}, assuming a single speed of 20~km~s$^{-1}$ and a single density of 2500~kg~m$^{-3}$.

However, the Meteoroid Engineering Model \citep[MEM;][]{moorhead20} offers a more complete model of the sporadic flux, including a speed and density distribution. MEM is typically used to assess the risk of penetration posed by meteoroid to spacecraft, and thus many programs will have already assessed the sporadic risk using this model. We therefore opt to compare shower fluxes with the sporadic flux as modeled by MEM~3 rather than \cite{grun85}, and suggest that this approach be considered for incorporation into the shower forecasting algorithm.

MEM is designed to compute the mass-limited flux relative to a spacecraft trajectory, but it is possible to use it to model the kinetic-energy-limited flux relative to the top of the Earth's atmosphere (which, from a meteoroid perspective, corresponds to an altitude of 100~km) with a carefully constructed input ``trajectory'' and some post-processing. Our so-called input trajectory consists of a series of state vectors with positions scattered randomly across the top of the Earth's atmosphere and velocity vectors that are aligned with the rotation of the Earth's surface. MEM's outputs then provide the flux of meteoroids with masses of at least 1~$\upmu$g at locations near the top of the Earth's atmosphere, $f_{\mu \textrm{g}}$.

We then re-weight this mass-limited flux to obtain the 105~J energy-limited flux. MEM partitions the flux by speed, direction, and bulk density. Of these quantities, only speed is relevant to the kinetic energy conversion; thus, we compute
\begin{align}
    f_\textrm{105~J} &= \sum_i f_{\mu \textrm{g}} \frac{g(m_\textrm{lim} = 2 \cdot \textrm{105 J}/v_i^2)}{g(1~\mu\textrm{g})}
\end{align}
where $v_i$ is the meteoroid speed relative to the top of the Earth's atmosphere in the $i$\textsuperscript{th} speed bin. The quantity $g(m)$ refers to Eq.~A3 of \cite{grun85}: while MEM's flux differs from that of \citeauthor{grun85}, it has the same mass-dependence and thus $g$ can be used to re-scale MEM's fluxes.

\subsection{Minimum ZHR or flux calculation}

The MEO's shower forecasting algorithm is outlined in \cite{moorhead19b} and is largely based on \cite{koschack90a}. The algorithm converts ZHR values to flux, taking the in-atmosphere speed and population or mass index into account. Rather than apply this process to our entire list of showers, many of which do not have measured population or mass indices, we have opted to invert it and determine the minimum ZHR at which a shower becomes significant as a function of speed and population index.

We begin by setting the flux equal to 5\% of the kinetic-energy-limited sporadic meteoroid flux at the top of the atmosphere: this is approximately equal to $f_\textrm{105 J}  = 0.0168$~km$^{-2}$~yr$^{-1}$. We then compute the corresponding flux of meteors that are brighter than magnitude 6.5:
\begin{align}
    f_{6.5} &= f_\textrm{105 J} \left( \frac{m_\textrm{105 J}}{m_{6.5}} \right)^{s-1}
    \label{eq:minf65}
\end{align}
where $m_\textrm{105 J}$ is the mass of a meteoroid that corresponds to a kinetic energy of 105~J, $m_{6.5}$ is the mass of a meteoroid whose magnitude is 6.5, and $s$ is the mass index and governs the steepness of the shower's mass distribution (see Eq.~\ref{eq:index}). Both masses depend on the speed of the shower meteoroids at the top of the atmosphere:
\begin{align}
    v^2_\textsc{toa} &= v_g^2 + v_\mathrm{esc}^2 (h = 100~\mathrm{km})
\end{align}
Once $v_\textsc{toa}$ has been determined, we can compute the relevant masses:
\begin{align}
    m_\textrm{105 J} &= \frac{2 \times \textrm{105 J}}{v^2_\textsc{toa}} \\[0.5em]
    m_{6.5} &= 10^{6.0473 - 0.4348 \times 6.5} ~ v_\textsc{toa}^{-4.25}
\end{align}
Note that we have introduced the terms $v_\textsc{toa}$ and $m_\textrm{105 J}$ to denote the speed of the shower meteoroids at the top of the atmosphere (100~km) and the mass that corresponds to a kinetic energy of 105~J at that speed. 

Finally, we convert to ZHR as follows:
\begin{align}
    \mathrm{ZHR} &= \frac{f_{6.5} \times 37\,200~\textrm{km}^2}{(13.1 r - 16.5)(r-1.3)^{0.748}}
    \label{eq:minzhr}
\end{align}
where $r$ is the population index. 

Equation~\ref{eq:minf65} can be used to assess the significance of radar showers, and Eq.~\ref{eq:minzhr} can be used to assess the significance of visually observed showers. Equation~\ref{eq:minf65} can also be modified to assess shower significance using any arbitrary mass limit; one need only replace $m_{6.5}$ with the desired limiting mass.

\section{Results}
\label{sec:list}

\subsection{Minimum required ZHR}
\label{sec:minzhr}

\begin{figure*}
    \centering
    \includegraphics[width=0.9\linewidth]{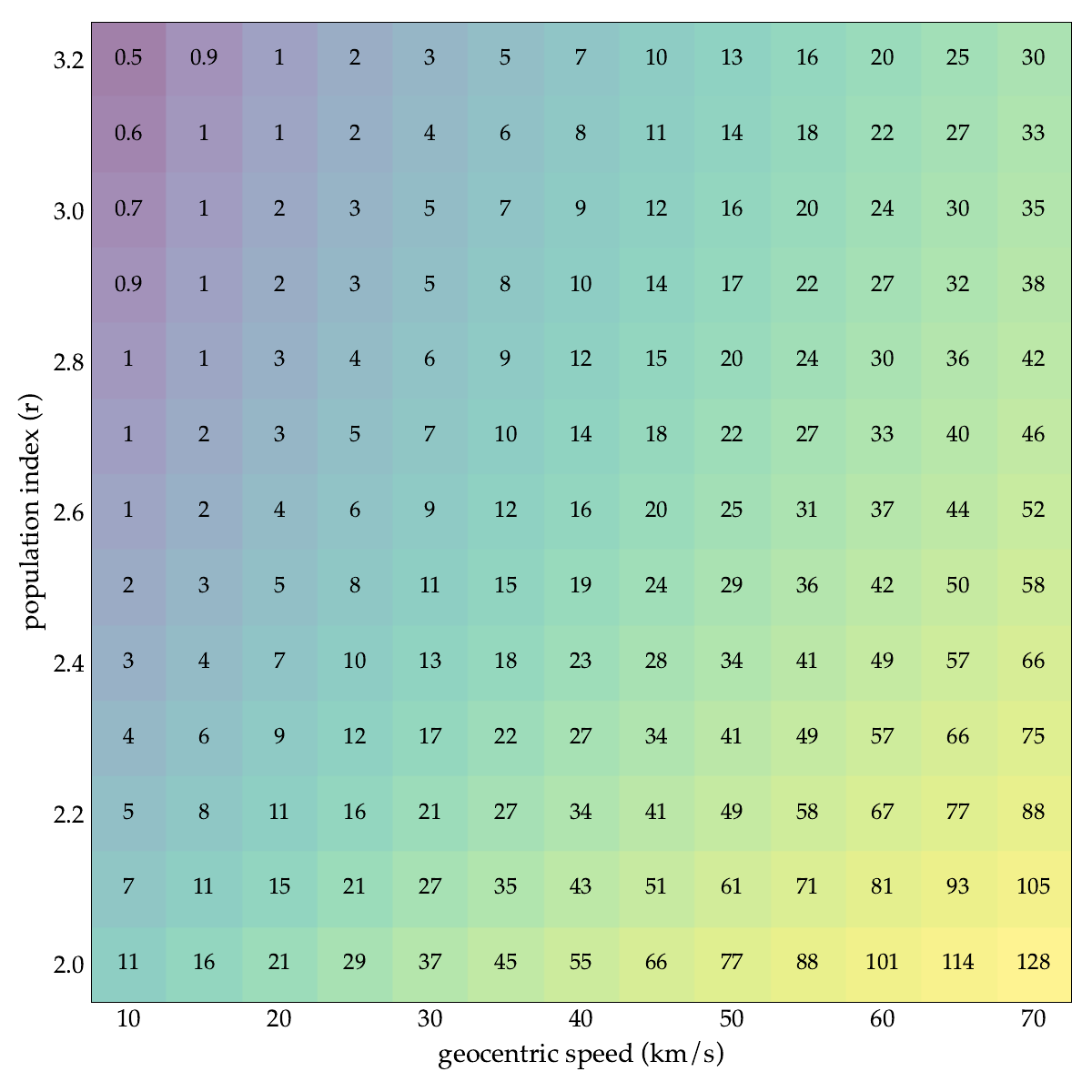}
    \caption{The minimum ZHR value at which a meteor shower contributes 5\% of the flux of meteoroids carrying a kinetic energy of at least 105~J in near-Earth space, as a function of population index, $r$, and geocentric speed, $v_g$.}
    \label{fig:zhrgrid}
\end{figure*}

Figure~\ref{fig:zhrgrid} provides the minimum ZHR for which a shower can be considered potentially hazardous according to our criterion as a function of population index, $r$, and the speed ``at infinity,'' $v_g$.  The quantity $v_g$ is also sometimes called the geocentric speed. We see that this ZHR threshold varies by more than two orders of magnitude over our grid; for showers that are both very slow (${v_g = 10}$~km~s$^{-1}$) and rich in small particles (${r = 3.2}$), a ZHR of just 0.5 can produce a 5\% enhancement in the flux. Slow meteoroids are disproportionately dim; if the stream also has fewer large particles than normal (i.e., a high value of $r$), a small number of bright meteors per hour is the tip of a vast ``iceberg'' containing many small, slow, but still hazardous particles.

On the other hand, a high-speed meteor shower with many large particles can be visually spectacular without posing much of a threat if its population index is low. For instance, the epsilon Geminids (EGE) are an exceptionally fast shower ($v_g \simeq 70$~km~s$^{-1}$) and the NASA All Sky Fireball Network \citep{cooke12} has detected a number of EGE fireballs. However, it has a relatively modest ZHR and low population index and, as a result, poses an utterly insignificant risk to spacecraft. It would require an outburst of at least several hundred EGE meteors to produce even a modest risk. Such showers have ``more bark than bite.''

\subsection{Minimum required magnitude-6.5-limited flux}

\cite{koschack90a} provide a framework for converting any magnitude- or mass-limited flux to ZHR and vice versa. However, when the flux is derived from observations of meteors that are far smaller than those visible to the naked eye, such an extrapolation can be misleading or inaccurate. For instance, the theta Coronae Borealids (TCBs) appear to produce a significant flux at radar sizes \citep{brown10,profilesreport}, but are not observed visually. Thus, the \emph{actual} ZHR is near zero, but the radar-derived ZHR can be greater than zero.

Furthermore, the meteor community is not in agreement on the relationship between magnitude and mass. This can cause further complications in the use of ZHR. For instance, if one researcher converts radar flux to ZHR assuming that brightness is proportional to mass \citep[e.g.,][]{cb04}, and then we convert ZHR back to energy-limited flux using \cite{koschack90a}, who assume that brightness is proportional to mass$^{0.92}$, the result would be incorrect.

\begin{figure*}
    \centering
    \includegraphics[width=0.9\linewidth]{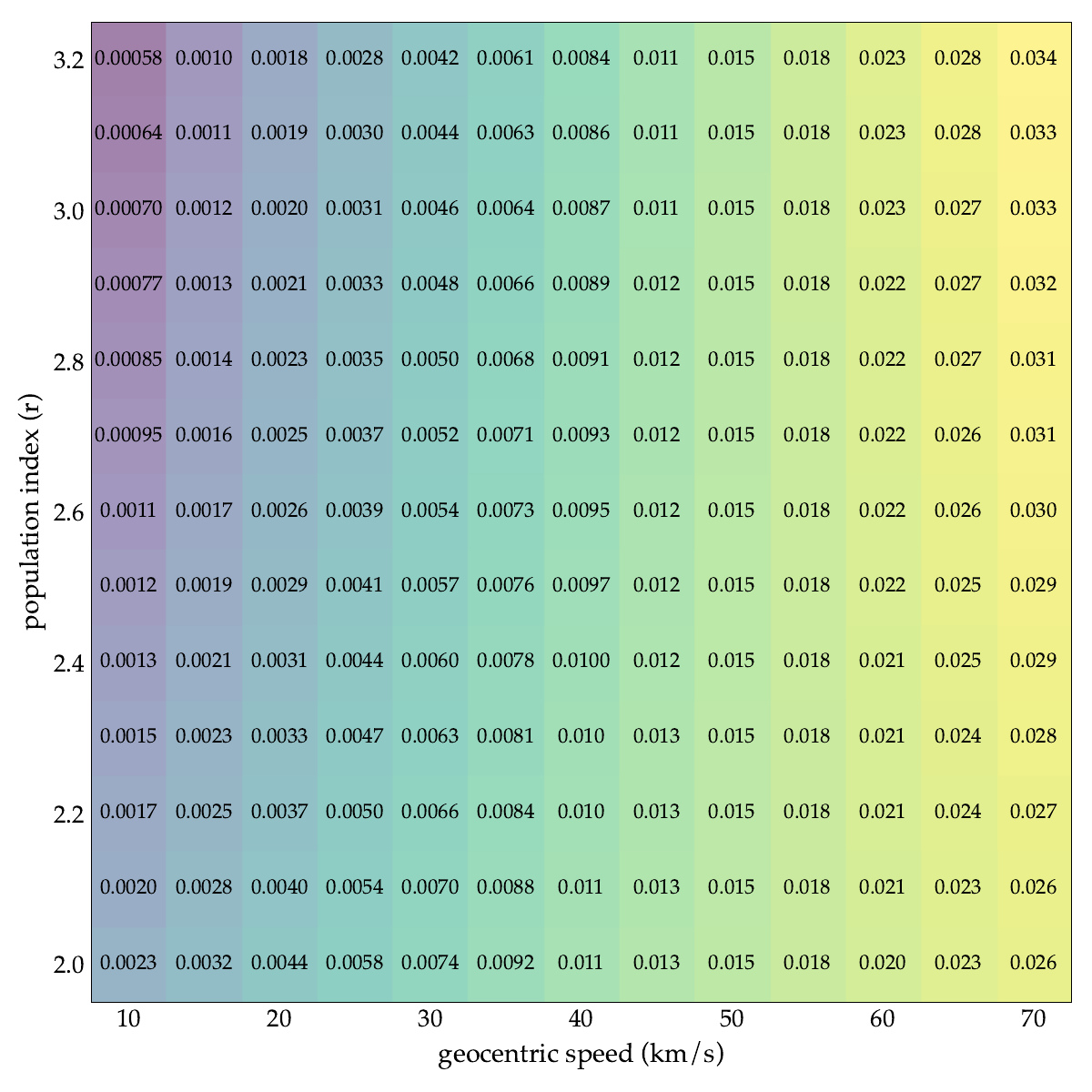}
    \caption{The minimum magnitude-6.5-limited flux, in units of km$^{-2}$~hr$^{-1}$, at which a meteor shower contributes 5\% of the flux of meteoroids carrying a kinetic energy of at least 105~J in near-Earth space, as a function of population index, $r$, and geocentric speed, $v_g$.}
    \label{fig:f65grid}
\end{figure*}

To minimize the impacts of incongruous mass-magnitude relations, we provide a second threshold grid in Fig.~\ref{fig:f65grid}. This grid depicts the minimum magnitude-6.5-limited flux, in units of km$^{-2}$~hr$^{-1}$, for which a shower can be considered potentially hazardous. This choice is based on the work of \cite{cb04}, who found that magnitude-6.5-limited fluxes were relatively insensitive to the choice of mass index, at least for CMOR and the Daytime Arietids. We expect this grid to be more useful to those working with radar meteor fluxes.

\subsection{Potentially hazardous meteor showers}
\label{sec:phms}

We will take a conservative approach in which we assume that the uncertainty in the population index, $r$, is large. For most showers, we will assume that there is no useful constraint on $r$ and that possible values range from 1.6 to 3.3. For those showers with published population indices, we assume an uncertainty of $\pm 0.4$, regardless of whether a smaller published uncertainty is available. This is motivated in part by the fact that the population and mass index could vary with particle size in a manner that cannot be captured by a single data set.

\subsubsection{Strong annual showers}

A handful of meteor showers not only satisfy our criterion for being potentially hazardous, but exceed the threshold by at least a factor of two in a typical year (see Table~\ref{tab:strong}). For instance, the Geminids (GEM) produce 10-20 times the level of activity needed to produce a 5\% enhancement over the sporadic flux. We label these showers ``strong annual showers.'' 

\begin{table*}
    \centering
\begin{tabular}{l@{\hskip 24pt}c@{\hskip 24pt}ccc@{\hskip 24pt}cc@{\hskip 24pt}cc} 
\hline \hline \addlinespace[2pt]
code & $v_g$ & $r_\text{min}$ & $r$ & $r_\text{max}$ &
    ZHR$_\text{req}$ & ZHR & $f_\text{req}$ & $f$ \\  \addlinespace[2pt]
\hline \addlinespace[2pt]
ARI  & 38.7 & 2.4 & 2.8 & 3.2 &   7 &  \og 30 & 7.7e-03 & \og 2.0e-02 \\ \addlinespace[2pt]
GEM  & 34.4 & 2.2 & 2.6 & 3.0 &   7 & \og 150 & 6.2e-03 & \og 5.7e-02 \\ \addlinespace[2pt]
PER  & 59.9 & 1.8 & 2.2 & 2.6 &  38 & \og 100 & 2.0e-02 & 5.4e-03 \\ \addlinespace[2pt]
QUA  & 41.1 & 1.7 & 2.1 & 2.5 &  20 & \og 110 & 1.0e-02 & \og 3.4e-02 \\ \addlinespace[2pt]
SDA  & 40.6 & 2.9 & 3.3 & 3.3 &   7 &  \yl 11 & 8.6e-03 & \og 4.5e-02 \\ \addlinespace[2pt]
XSA$^*$  & 25.4 & 1.6 &     & 3.3 &   2 &     & 2.8e-03 & \og 5.7e-03 \\ \addlinespace[2pt]
\hline \addlinespace[2pt]
\end{tabular}
\caption{Significance thresholds and observed activity levels for established meteor showers that typically produce at least twice the required ZHR \emph{or} the required magnitude-6.5-limited flux to be considered potentially hazardous; we classify these as ``strong annual showers.'' Quantities that exceed the threshold by a factor of two are highlighted in green, and those that exceed the threshold by less than a factor of two are highlighted in yellow. \\[4pt]
We provide the shower's three-letter abbreviation code and geocentric speed in the first two columns. The next three columns provide the minimum assumed population index, the nominal population index (if known), and the maximum assumed population index. We then provide the minimum ZHR required for our hazard threshold for any population index in the assumed range, followed by the peak ZHR (if known). Finally, we provide the minimum flux required for our hazard threshold for any population index in the assumed range, followed by the peak flux (if known). \\[4pt]
Showers marked with an asterisk may appear strong due in part to outdated or missing population index measurements, and would benefit from further study.
}
\label{tab:strong}
\end{table*}

Table~\ref{tab:strong} includes one relatively obscure shower: the Daytime xi Sagittariids (XSA). This shower produced the 13\textsuperscript{th} largest peak flux measured by \cite{profilesreport}. It also has a fairly low speed and unmeasured population index. When we assume that the population index could be as steep as 3.3, as we have done here, the minimum flux required for the shower to meet our hazard threshold is only 0.0028~km$^{-2}$~hr$^{-1}$, the lowest in the table. However, if the true population index were, for example, ${r = 2.0}$, we see from Fig.~\ref{fig:f65grid} that the threshold flux would be 0.0058~km$^{-2}$~hr$^{-1}$, which is slightly higher than the measured peak XSA flux. This would not only bump it off the strong showers list, but out of the hazardous shower category entirely. Thus, the XSAs, and their population index in particular, are a prime candidate for future study.

\subsubsection{Moderate annual showers}

Eleven additional showers exceed the hazard threshold, but not by a factor of two (Table~\ref{tab:moderate}). We label these showers ``moderate annual showers.'' Many showers in Table~\ref{tab:moderate} lack reliable population index measurements: for instance, the SSG index is based on a 30-year-old study of visual observations \citep{jenniskens94}. Thus, most showers in our list of moderate annual showers, with the possible exception of the eta Aquariids (ETA), Daytime Sextantids (DSX) and Ursids (URS), merit further study in order to firmly establish their status as potentially hazardous.

\begin{table*}
    \centering
\begin{tabular}{l@{\hskip 24pt}c@{\hskip 24pt}ccc@{\hskip 24pt}cc@{\hskip 24pt}cc} 
\hline \hline \addlinespace[2pt]
code & $v_g$ & $r_\text{min}$ & $r$ & $r_\text{max}$ &
    ZHR$_\text{req}$ & ZHR & $f_\text{req}$ & $f$ \\ \addlinespace[2pt]
\hline \addlinespace[2pt]
AUD$^{*,\dagger}$  & 19.2 & 1.6 &     & 3.3 &   1 &   \yl 1.7 & 1.5e-03 &         \\ \addlinespace[2pt]
CAP  & 22.6 & 2.1 & 2.5 & 2.9 &   3 &  \yl 5 & 2.7e-03 & \yl 4.2e-03 \\ \addlinespace[2pt]
DSX  & 31.8 & 1.6 & 1.8 & 2.2 &  24 &     & 7.2e-03 & \yl 1.3e-02 \\ \addlinespace[2pt]
ETA  & 65.7 & 2.0 & 2.4 & 2.8 &  37 &  \yl 50 & 2.3e-02 & 1.9e-02 \\ \addlinespace[2pt]
GDR  & 27.4 & 2.6 & 3.0 & 3.3 &   3 &   \yl 5 & 3.3e-03 &         \\ \addlinespace[2pt]
HVI$^{*,\dagger}$  & 18.8 & 1.6 &     & 3.3 &   1 &  \yl 1.6 & 1.5e-03 &         \\ \addlinespace[2pt]
SSG  & 23.7 & 2.5 & 2.9 & 3.3 &   2 &  \yl 2 & 2.4e-03 &         \\ \addlinespace[2pt]
STA  & 26.8 & 1.9 & 2.3 & 2.7 &   6 &  \yl 7 & 4.2e-03 & \yl 4.3e-03 \\ \addlinespace[2pt]
SZC$^*$  & 37.7 & 1.6 &     & 3.3 &   6 &     & 7.1e-03 & \yl 1.3e-02 \\ \addlinespace[2pt]
TCB$^*$  & 38.2 & 1.6 &     & 3.3 &   6 &     & 7.4e-03 & \yl 7.7e-03 \\ \addlinespace[2pt]
URS  & 34.2 & 2.4 & 2.8 & 3.2 &   5 &  \yl 10 & 5.8e-03 & \yl 8.5e-03 \\ \addlinespace[2pt]
\end{tabular}
\caption{Significance thresholds and observed activity levels for established meteor showers that typically produce either the required ZHR or the required magnitude-6.5-limited flux to be considered potentially hazardous; we classify these as ``moderate annual showers.'' The quantities that exceed the threshold are highlighted in yellow. Showers marked with an asterisk may appear strong due in part to outdated or missing population index measurements; those marked with a dagger have ZHR values estimated from the \cite{vr} video rate.}
\label{tab:moderate}
\end{table*}

Readers may notice that Table~\ref{tab:moderate} includes the Southern Taurids (STA) but not the Northern Taurids. This is because the estimated NTA ZHR is 5, slightly lower than that of the STAs. We were also not successful in fitting the radar activity profile of the NTAs in \cite{profilesreport}, which may indicate that the northern branch is also slightly weaker at radar magnitudes. However, the two branches have a long overlap in activity and have similar radiants. It may be worth considering them a single, potentially hazardous complex.

Four showers in Table~\ref{tab:moderate} do not have population index estimates available from the IMO meteor shower calendar or from past forecasts. If their true population index is low, these showers may drop off the list of moderate showers in the same manner as the XSAs. Two of these showers (AUD and HVI) also lack ZHR and flux measurements, and their inclusion is based on the ZHR estimated from the \cite{vr} video rates. Thus, these showers are particularly in need of further study.

\subsubsection{Outbursting showers}

Table \ref{tab:variable} lists a number of showers that have produced at least one outburst that satisfies our criterion. This list is not exhaustive: the scarcity of data on some of these outbursts -- such as the 1927 June Bo\"{o}tids (JBO) -- suggests the possibility that additional outbursts might lurk in obscure publications or have gone completely unobserved. We list only the strongest (or, in some cases, a combination of the strongest and most reliably measured) outburst for each shower. The Leonids (LEO) have exceeded our hazard threshold many times, while the beta Hydrusids (BHY) have been observed only once.

\begin{table*}
    \centering
\begin{tabular}{l@{\hskip 24pt}c@{\hskip 24pt}ccc@{\hskip 24pt}ccll} 
\hline \hline \addlinespace[2pt]
code & $v_g$ & $r_\text{min}$ & $r$ & $r_\text{max}$ &
    ZHR$_\text{req}$ & ZHR$_\text{max}$ & year & ZHR$_\text{max}$ reference \\ \addlinespace[2pt]
\hline \addlinespace[2pt]
AMO & 63.0 & 2.0 & 2.4 & 2.8 &  34 & 2000 & 1935 & \protect\cite{kresak58}     \\
AND & 17.4 & 2.3 & 2.7 & 3.1 &   1 & 3600 & 1885 & \protect\cite{denning85}    \\
ARD$^\ddagger$ & 10.8 & 2.1 & 2.5 & 2.9 &   1 &  80 & 2021 & 
    \cite{jenniskens21} \\
AUR  & 66.0 & 2.1 & 2.5 & 2.9 &  34 & 200 & 1986, 1994 & \cite{jenniskens08}  \\ 
BHY & 24.0 & 2.2 & 2.1 & 3.0 &   3 & 80 & 1985 & \protect\cite{jenniskens06}    \\
CAM$^\ddagger$ & 13.9 & 2.8 & 3.2 & 3.3 &   1 & 20 & 2014 & \protect\cite{cb16} \\
DRA & 19.4 & 2.2 & 2.6 & 3.0 &   2 & 9000 & 2012 & \cite{ye14}   \\
JBO & 14.2 & 1.8 & 2.2 & 2.6 &   3 & 250 & 1998 & \cite{jenniskens06}   \\
LEO & 70.0 & 2.1 & 2.5 & 2.9 &  39 & 150,000 & 1966 & \cite{milon67,brown97} \\
LYR & 47.0 & 1.7 & 2.1 & 2.5 &  27 & 90-100 & 1982 & \cite{mcleod82} \\
ORI & 66.2 & 1.7 & 2.1 & 2.5 &  52 & 80 & 2007 & \cite{arlt08} \\ \addlinespace[2pt]
PHO & 11.3 & 2.4 & 2.8 & 3.2 &   1 & 100 & 1956 & \cite{venter57} \\
PPU & 15.4 & 1.6 & 2.0 & 2.4 &   5 & 40 & 1977 & \cite{shao77} \\
TAH & 16.1 & 1.8 & 2.2 & 2.6 &   3 & 40-50 & 2022 & \cite{weiland22} \\
XDR & 35.8 & 2.3 & 2.7 & 3.1 &   7 & 20 & 1996 & \cite{langbroek96}
\end{tabular}
\caption{Significance thresholds and strongest outbursts for a selection of variable showers. This list includes several showers with annual activity, such as the Leonids (LEO), April Lyrids (LYR), and Orionids (ORI). These showers do not meet our hazard threshold in a typical year, but can far exceed it when in outburst. Showers marked with a double dagger are not in the IAU's list of established showers.
}
\label{tab:variable}
\end{table*}

Although we have separated variable showers from annual showers in our classification, there is no hard division between the two. For instance, the Perseids (PER) can vary in intensity from year to year; in the early 1990s, the shower attained ZHR levels between 200 and 300. Here, we have categorized showers based on their typical activity level, and relegate them to Table~\ref{tab:variable} only if their annual activity does not meet our threshold.

Some of the outbursts in Table~\ref{tab:variable} exceeded the hazard threshold by many orders of magnitude. For instance, the 2012 Draconids, 1966 Leonids, and 1885 Andromedids each increased the 105~J meteoroid flux by more than two orders of magnitude. The Draconids in particular pose a significant risk to spacecraft and can ``fly under the radar.'' A significant Draconid outburst with a peak ZHR of 1250 occurred in 1999 and, at the time, went completely undetected. This outburst was uncovered two decades later when \cite{egal19} noticed that their model consistently predicted activity in 1999, and subsequently located evidence of the outburst in data gathered during a calibration survey prior to the 1999 Leonids.

In one case, we manually adjusted the nominal estimate of the population index. The 2021 ARD outburst was measured both visually \citep{jenniskens21} and via radar \citep{janches23}. Both works attempted to measure the population index and produced estimates that are wildly disparate: \cite{jenniskens21} reports ${r=4.2}$, while \cite{janches23} reports ${s = 1.69}$, which corresponds to ${r=1.9}$. It may be that the ARDs deviate from a power law; however, we find that a ZHR of 80 \citep{jenniskens21} can be brought into agreement with a magnitude-9-limited flux of 0.36~km$^{-2}$~hr$^{-1}$ if the population index is assumed to have a more typical value of 2.5. We therefore adopt ${r=2.5}$ for the purposes of extrapolating the ARD flux to our 105-J threshold.

\subsection{Observational bias and completeness}

We were not able to assess the entire list of established showers published by the IAU Meteor Data Center. Our literature search yielded no activity measurements for 36 established showers. For three additional showers, the only activity measurement available was non-detection reported by \cite{vr}. Thus, we cannot claim that our lists of potentially hazardous showers are complete. Furthermore, our list contains five showers with unmeasured population indices; we therefore also cannot claim that our list is limited to potentially hazardous showers.

In this section, we attempt to probe the degree to which our lists may be biased by testing the distribution of showers for Sun-centered ecliptic symmetry.
The sporadic meteor complex is believed to arise primarily from comets \citep{jones01,wiegert09,nesvorny11,pokorny14}; similarly, most known meteor shower parent bodies are comets, aside from a couple of comet-like asteroids (2003 EH1 and 3200 Phaethon). For this reason, we expect the distribution of meteor shower radiants to resemble that of the sporadic complex.
Although the meteoroid environment is certainly not isotropic in its directionality, the distribution of sporadic radiants is both vertically and horizontally symmetric across the apex direction when viewed in a Sun-centered coordinate system \citep[e.g.,][]{cb08}. 

There are, on the other hand, observational biases that can produce artificial asymmetries. Optical networks are unable to detect daytime meteor showers; radar networks can detect meteors day and night, but daytime meteor detection can be suppressed by Faraday rotation \citep{ceplecha98} and enhanced noise due to solar radio emission. Meteors are equally visible from the northern and southern hemispheres, but 87\% of the world's population lives in the northern hemisphere, possibly biasing shower detections towards those with northern declinations. Furthermore, the radar flux data used in this analysis is from CMOR, which is located in Ontario and cannot observe showers with declinations ${\lesssim -40^\circ}$ \citep{cb06}. \cite{vr} reported a similar bias and claimed that their survey was complete to a declination of -25$^\circ$.

\begin{figure*}
    \centering
    \includegraphics{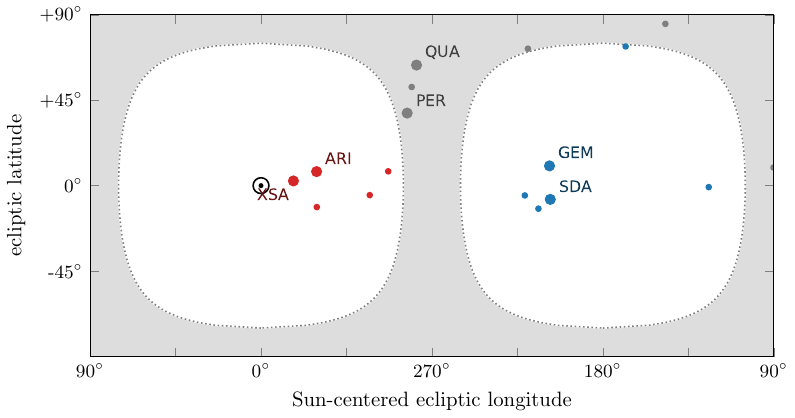}
	\caption{Sun-centered ecliptic radiants of strong annual showers (large, annotated points) and moderate annual showers (small points). The angular position of the Sun is represented by the ``$\odot$'' symbol. ``Daytime'' showers appear in red, while nighttime showers are in blue. Showers in the gray region have solar elongation angles between 75$^\circ$ and 105$^\circ$ and are excluded from our comparison.}
	\label{fig:sun}
\end{figure*}

Figure~\ref{fig:sun} presents the Sun-centered ecliptic radiants of showers listed in Tables~\ref{tab:strong} and \ref{tab:moderate}. We see that there are equal numbers of daytime and nighttime meteor showers: there are exactly two strong annual showers in each category. Moderate annual showers are also evenly divided, with three daytime and four nighttime showers. We exclude those showers that lie near the dividing line; specifically, we ignore showers with solar elongation angles between 75 and $105^\circ$. Our inclusion of the \cite{profilesreport} peak radar fluxes appears to have resulted in a selection that is not obviously biased towards nighttime showers.

The north-south distribution of shower radiants, however, is not even. Here, we ignore those showers that lie within $15^\circ$ of the ecliptic. Only two strong annual showers -- the Quadrantids and the Perseids -- remain; we are unable to comment on the north-south distribution on the basis of two showers. However, 4 moderate showers lie more than $15^\circ$ from the ecliptic, none of which lie south of the ecliptic. If we consider both groups, 0 out of 6 showers is a statistically significant deviation from symmetry (${p=0.03}$ for a two-tailed test), and suggests that our list of annual showers may be biased towards the northern hemisphere \citep[see also][]{neslusan14,jenniskens18}.

\begin{figure*}
    \centering
    \includegraphics{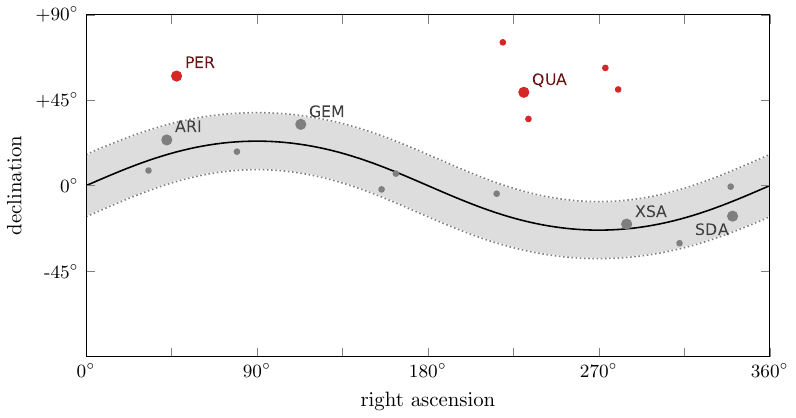}
	\caption{Radiants of strong (large, annotated points) and moderate annual showers (small points) in equatorial coordinates. The sinusoidal black line represents the ecliptic, and the gray band spans declinations that are within 15$^\circ$ of the ecliptic. Outside of this exclusion region, northern showers appear in red; none of the showers in our list lie south of the gray band.}
	\label{fig:ns}
\end{figure*}

Resolving this possible bias will involve at least two avenues of research. First, the list of current candidates needs further characterization. Some of these showers could drop off the list if their population indices are measured and determined to be much lower than our assumed upper bound of 3.3. 

Second, our list would benefit from the inclusion of southern hemisphere radar fluxes. Most of the showers in our lists are included on the basis of their peak radar flux. \cite{pokorny17} recently surveyed meteor showers detected by the Southern Argentina Agile MEteor Radar (SAAMER), another patrol radar located at a latitude of approximately 54$^\circ$~S, and reported a large number of previously unknown showers with declinations below $-40^\circ$. The SAAMER team has also recently started to publish meteor fluxes \citep{bruzzone21,janches23}. So far, flux values have been published only for the 2020 A Carinid (which did not meet our hazard threshold) and 2021 Arid outbursts. If more fluxes become available, SAAMER could very well help to identify additional hazardous showers at southern declinations. However, we would not have been able to fit many of the 38 showers we characterized in \cite{profilesreport} without combining 20 years of observations; we may have to wait a similar length of time before we can do the same with southern showers.

\section{Conclusion}

We have proposed a quantitative criterion for determining whether a meteor shower poses a potential risk to spacecraft. We require that such a shower elevates the hazardous meteoroid flux onto a fully exposed spacecraft surface in low Earth orbit by at least a modest amount (5\% over the sporadic rate). We evaluate the flux contribution at a limiting kinetic energy of 105~J, which is representative of the energy required to damage relatively delicate spacecraft components.

Such showers pose a risk only to spacecraft; observers on the ground are well-protected by the atmosphere. Furthermore, the required 5\% increase in flux is rather modest: we do not suggest that spacecraft need to mitigate for this level of shower activity. Instead, we use our criterion as a threshold for investigation: showers that meet it should be included in environment descriptions and meteor shower forecasts. 

Determining whether a meteor shower meets this threshold requires some measurement of the shower's peak activity, population index, and speed.  We have collected activity measurements or estimates for 74 of the 110 meteor showers listed as ``established'' in the IAU Meteor Data Center (as of 25 April 2023), and have evaluated whether these shower could meet our hazard threshold for any reasonable choice of index. We found that six showers not only meet our threshold, but exceed it by at least a factor of two; these showers are the Daytime Arietids (ARI), Geminids (GEM), Perseids (PER), Quadrantids (QUA), Southern delta Aquariids (SDA), and Daytime xi Sagittariids (XSA). An additional 11 showers meet the threshold but do not double it. We also provide examples of 14 variable showers that do not meet the threshold in a typical year, but can exceed it, sometimes by orders of magnitude, when in outburst.

Our list of hazardous meteor showers is not complete; we applied our criterion only to showers for which peak ZHR or flux values were readily available. Neither is it limited to those showers that are truly hazardous, as it includes five showers that have unmeasured population indices and/or ZHR or flux values. As a result, the list includes not only showers like the Geminids that clearly produce a significant hazardous flux, but also showers like the Daytime xi-Sagittariids (XSA) that \emph{may} be hazardous but require further study. We hope that our colleagues in the meteor astronomy community use our criterion in conjunction with new meteor observations to continually assess and re-assess the significance of meteor showers for spacecraft safety. In order to enable such assessments, we have provided tables of the ZHR (Fig.~\ref{fig:zhrgrid}) or flux (Fig.~\ref{fig:f65grid}) at which a shower with a given speed and population index meets the threshold.

\section{Acknowledgments}

This work was supported in part by NASA Cooperative Agreements 80NSSC21M0073 and 80NSSC24M0060 and by the Natural Sciences and Engineering Research Council of Canada.

AVM would like to thank Danielle Moser for providing a number of useful references on meteor shower outbursts.

\bibliography{local}

\appendix

\section{Master list}

Tables~\ref{tab:master1}-\ref{tab:master3} present the data we collected for each established meteor shower plus six additional showers of special interest. We provide the population index; the peak ZHR, video rate, or radar flux; and the source(s) from which we collected this information. If none of these quantities could be obtained for an established shower, we list only its three-letter abbreviation.

\begin{table*}
\begin{tabular}{lccccl} \hline \hline
code & $r$ & ZHR & video rate & radar flux 
    & additional source(s) \\ 
    &&&& (km$^{-2}$ hr$^{-1}$) \\ \hline
AAN &  & 1 &  & 0.00947 & \cite{molau09feb} \\
AHY & 2.8 & & 1.5 &  & \cite{jenniskens94} \\
ALA &  &  &  &  \\
AMO & 2.4 & var (2000) &  &  & \cite{cal23,kresak58} \\
AND & 2.7 & var (3600) & 0.9 &  & \cite{wiegert13,denning85} \\
APS &  &  &  &  \\
ARC &  &  &  &  \\
ARD$^\dagger$ & 2.5 & var (80) &  &  & \cite{jenniskens21,janches23} \\
ARI & 2.8 & 30 &  & 0.059 & \cite{cal23} \\
AUD &  & & 1.1 &  & \\
AUM$^\dagger$ &  &  &  & 0.00624 \\
AUR & 2.5 & var (200) &  &  & \cite{cal23,jenniskens08} \\
AVB &  &  & 0 &  \\
BEQ &  &  &  & 0.0127 \\
BHY & 2.1 & var (80) &  &  & \cite{ams,lunsford09} \\
BTA & 2.2 &  &  & 0.00777 & \cite{dewsnap21} \\
CAM$^\dagger$ & 3.2 & var (20) &  &  & \cite{cb16} \\
CAN &  & & 2.3 &  &  \\
CAP & 2.5 & 5 & 5.2 & 0.0127 & \cite{cal23} \\
COM & 3 & 3 & 2.4 &  & \cite{cal23} \\
COR &  &  & 0 &  \\
CTA &  &  &  &  \\
DAD &  & & 1.7 & & \\
DKD &  &  &  &  \\
DLT &  &  &  & 0.0151 \\
DPC &  &  &  &  \\
DRA & 2.6 & var (9000) &  &  & \cite{cal23,ye14} \\
DSV &  &  &  &  \\
DSX & 1.8 &  &  & 0.0399 & \cite{kipreos22} \\
EAU &  &  &  &  \\
ECV &  &  &  &  \\
EGE & 3 & 3 & 2.8 &  & \cite{cal23} \\
EHY &  &  &  &  \\
ELY & 3 & 3 & 2.8 &  & \cite{cal23} \\
EPG &  &  &  & 0.0112 \\
EPR &  &  &  &  \\
ERI & 3 & 3 & 4.9 &  & \cite{cal23} \\
ETA & 2.4 & 50 & 54 & 0.0583 & \cite{cal23} \\
EVI &  &  &  &  \\
\end{tabular}
\caption{Population index ($r$) and activity measurements used for the meteor showers considered in this report; in this table we list showers with three-letter codes beginning with letters A-E. A dagger indicates that the shower is not in the Meteor Data Center's list of established showers. The listed sources are those for $r$ and ZHR; all video rates are from \cite{vr} and all radar fluxes are from \cite{profilesreport}.}
\label{tab:master1}
\end{table*}

\begin{table*}
\begin{tabular}{lccccl} \hline \hline
code & $r$ & ZHR & video rate & radar flux 
    & additional source(s) \\ 
    &&&& (km$^{-2}$ hr$^{-1}$) \\ \hline
FAN &  &  &  &  \\
FED &  &  &  &  \\
FEV &  &  &  &  \\
GDR & 3 & 5 & 1.6 &  & \cite{cal23} \\
GEM & 2.6 & 150 & 46 & 0.172 & \cite{cal23} \\
GUM & 3 & 3 &  & 0.00495 & \cite{cal23} \\
HVI &  & & 1.6 &  & \\
HYD & 3 & 7 & 4.4 &  & \cite{cal23} \\
JBO & 2.2 & var (250) &  &  & \cite{cal23,jenniskens06} \\
JIP &  &  &  &  \\
JLE &  & 1 & 0.6 & 0.00935 \\
JMC &  &  &  &  \\
JPE & 3 & 5 & 1.6 &  & \cite{cal23} \\
JRC &  &  &  &  \\
JXA &  &  &  &  \\
KCG & 2.2 & 2.3 &  &  & \cite{jenniskens94} \\
KLE &  &  &  & 0.011 \\
KSE &  &  &  &  \\
KUM &  &  &  &  \\
LBO &  &  &  & 0.0171 \\
LDR$^\dagger$ &  &  &  & 0.0044 \\
LEO & 2.5 & var (150k) &  &  & \cite{milon67,brown97} \\
LLY$^\dagger$ &  &  &  & 0.0083 \\
LMI & 3 & 2 & 4.2 & 0.0056 & \cite{cal23} \\
LUM &  &  &  &  \\
LYR & 2.1 & var (100) & 13 & 0.0136 & \cite{cal23,mcleod82} \\
MKA &  &  &  &  \\
MON & 3 & 3 & 2.3 &  & \cite{cal23} \\
NCC & 1.75 & 8.6 &  &  & \cite{sokolova22} \\
NDA & 3.3 & 1 &  &  & \cite{jenniskens94} \\
NIA &  &  &  &  \\
NID$^\dagger$ &  &  &  & 0.006 \\
NOC &  &  &  & 0.016 \\
NOO & 3 & 3 & 3.2 & 0.0185 & \cite{cal23} \\
NTA & 2.3 & 5 & 4.1 &  & \cite{cal23} \\
NUE &  &  & 5.1 &  & \\
NZC & 2.7 & 4.7 &  & 0.0168 & \cite{holman12} \\
\end{tabular}
\caption{Population index ($r$) and activity measurements used for meteor showers with three-letter codes beginning with letters F-N.}
\label{tab:master2}
\end{table*}

\begin{table*}
\begin{tabular}{lccccl} \hline \hline
code & $r$ & ZHR & video rate & radar flux 
    & additional source(s) \\ 
    &&&& (km$^{-2}$ hr$^{-1}$) \\ \hline
OCC &  &  & 0 &  \\
OCE &  &  &  & 0.0133 \\
OCT & 2.5 & 5 & 2 &  & \cite{cal23} \\
OCU &  & & 2.5 &  & \\
OER &  & & 1.2 &  & \\
OHY &  &  &  &  \\
ORI & 2.1 & var (80) & 68 & 0.027 & \cite{arlt08} \\
ORS &  &  &  &  \\
OSE &  &  &  &  \\
PAU & 3.2 & 5 & 0 &  & \cite{cal23} \\
PCA &  &  & 0 & 0.0231 \\
PER & 2.2 & 100 & 68 & 0.0162 & \cite{cal23} \\
PHO & 2.8 & var (100) &  &  & \cite{cal23,venter57} \\
PPS &  &  &  &  \\
PPU & 2 & var (40) &  &  & \cite{ams,shao77} \\
PSU &  & & 1.8 &  \\
QUA & 2.1 & 110 & 23 & 0.1016 & \cite{cal23} \\
RPU &  &  &  &  \\
SCC & 1.75 & & 0.9 &  & \cite{sokolova22} \\
SDA & 3.3 & 11.4 & 19 & 0.1347 & \cite{jenniskens94} \\
SLD &  &  &  &  \\
SMA &  &  &  & 0.0108 \\
SPE & 3 & 5 & 2.6 &  & \cite{cal23} \\
SSE &  &  &  & 0.0118 \\
SSG & 2.9 & 2.4 &  &  & \cite{jenniskens94} \\
STA & 2.3 & 7 & 5.7 & 0.0128 & \cite{cal23} \\
SZC &  &  & 0 & 0.0395 \\
TAH & 2.2 & var (50) &  &  & \cite{ams,weiland22} \\
TCB &  &  &  & 0.023 \\
THA &  &  &  &  \\
URS & 2.8 & 10 & 7 & 0.0255 & \cite{cal23} \\
XCB &  &  & 1.2 & 0.0127 &  \\
XDR & 2.7 & var (20) &  &  & \cite{langbroek96} \\
XHE &  &  &  &  \\
XRI &  &  &  &  \\
XSA &  &  &  & 0.0172 \\
XUM &  &  &  &  \\
XVI &  &  &  &  \\
ZCA &  &  &  &  \\
ZPE & 1.9 &  &  & 0.0138 & \cite{dewsnap21} \\
\end{tabular}
\caption{Population index ($r$) and activity measurements used for meteor showers with three-letter codes beginning with letters O-Z.}
\label{tab:master3}
\end{table*}

\end{document}